\documentclass[preprint,12pt]{elsarticle}
\usepackage{graphicx}
\usepackage{hyperref}
\usepackage{mathrsfs}
\usepackage[english]{babel}
\usepackage{amsmath, amssymb, bm}
\usepackage{cleveref}
\usepackage[a4paper, margin=1in, centering]{geometry}
\usepackage[autostyle, english = american]{csquotes}
\MakeOuterQuote{"}

\date{\today}
\newcommand{\be}{\begin{eqnarray}}
\newcommand{\ee}{\end{eqnarray}}

\usepackage{xcolor}

\journal{Physics Letters B}

\begin{document}

\begin{frontmatter}




\title{A Type-I Seesaw Framework with Non-Holomorphic Modular Symmetry}






\author[1]{Cheshta}
\author[2]{Priya}
\author[1]{Suneel Dutt}
\author[2]{B. C. Chauhan}

\address[1]{Department of Physics, Dr. B.R. Ambedkar National Institute of Technology, Jalandhar, Punjab 144008, India}

\address[2]{Department of Physics and Astronomical Science, Central University of Himachal Pradesh, Dharamshala, Himachal Pradesh 176215, India}

\begin{abstract}
We study neutrino mass generation within the framework of non-holomorphic modular symmetry proposed by Qu and Ding. In this formalism, neutrino masses are generated via the Type-I seesaw mechanism, where the Yukawa couplings depend on non-holomorphic modular forms. The viability of the model is examined through a $\chi^2$ analysis using current neutrino oscillation data. The $\chi^2_{min}$ value is found to be $7.06$ for normal hierarchy(NH). All neutrino oscillation parameters are consistent within their $1\sigma$ allowed ranges, except the atmospheric mixing angle $\sin^2\theta_{23}$, which is predicted to lie in the second octant. The Dirac CP-violating phase($\delta_{CP}$) is constrained to the first and fourth quadrants, indicating relatively weak CP violation. These predictions can be tested in future long-baseline neutrino oscillation experiments. The sum of neutrino masses is compatible with the stringent bound proposed by the DESI experiment. However, the inverted hierarchy(IH) is not viable in this model, as the predicted value of $\chi^2_{min}$ exceeds 100, and the mixing angles $\sin^2\theta_{12}$ and $\sin^2\theta_{23}$ lie outside the $3 \sigma$ allowed ranges.
\end{abstract}




\end{frontmatter}




\section{Introduction} 
\label{introduction}
The Standard Model (SM) describes neutrinos as massless left-handed Weyl fermions. However, experimental observations from neutrino oscillation experiments such as the Super-Kamioka Neutrino Detection Experiment (Super-Kamiokande)~\cite{Super-Kamiokande:1998kpq}, the Sudbury Neutrino Observatory (SNO)~\cite{SNO:2002tuh}, and the Kamioka Liquid Scintillator Antineutrino Detector (KamLAND)~\cite{KamLAND:2002uet} demonstrate that neutrinos possess non-zero masses. This provides strong experimental motivation for physics beyond the Standard Model (BSM). Precision measurements from reactor and accelerator-based neutrino oscillation experiments have established all three mixing angles and the mass-squared differences with high accuracy. Despite this remarkable progress, the underlying mechanism responsible for neutrino mass generation and the fundamental nature of neutrinos remain open questions, thereby motivating extensions of the SM.

In particular, the Type-I seesaw mechanism arises from the introduction of heavy fermionic fields that are singlets under the Standard Model gauge group, namely $SU(2)_L$ singlet fermions, which generate light Majorana neutrino masses after electroweak symmetry breaking~\cite{Minkowski:1977sc}. In contrast, the Type-II seesaw mechanism is realized by introducing an $SU(2)_L$ scalar triplet with hypercharge $Y = 1$, whose vacuum expectation value directly contributes to the neutrino mass matrix~\cite{Mohapatra:1979ia}. The Type-III seesaw mechanism involves the addition of fermionic fields transforming as triplets under $SU(2)_L$ with zero hypercharge, which similarly induce small neutrino masses through their heavy Majorana mass terms~\cite{Foot:1988aq}.
 
After the precise measurement of the reactor mixing angle 
$\sin^2\theta_{13}$ by the Daya Bay~\cite{Leitner:2017jco}, RENO~\cite{RENO:2012mkc}, and Double Chooz~\cite{DoubleChooz:2011ymz} experiments, all neutrino oscillation parameters have been determined with good accuracy. The current global best-fit values and allowed ranges of these parameters are summarized in the NuFIT~6.0 analysis and are presented in Table~\ref{tab:neutrino_data}~\cite{Esteban:2024eli}. Despite this significant experimental progress, several fundamental questions remain open. In particular, the nature of neutrinos is still unknown, namely whether neutrinos are Dirac or Majorana particles. Neutrinoless double beta decay experiments aim to address this issue by testing lepton number violation~\cite{Priya:2025avk,Ankush:2023pax,Schechter:1981bd,Wolfenstein:1981rk}.

Several experiments are currently operating or planned for this purpose, including KamLAND-Zen~\cite{KamLAND-Zen:2024eml}, nEXO~\cite{nEXO:2021ujk}, CUORE~\cite{CUORE:2019yfd,CUORE:2020boz}, and LEGEND~\cite{LEGEND:2017cdu}. At present, KamLAND-Zen provides the most stringent experimental limit on neutrinoless double beta decay, searching for the decay in $^{136}\mathrm{Xe}$-doped liquid scintillator. It has established a lifetime limit of $\tau \geq 1.07$ × $10^{26}$ years at the 90\% confidence level (CL). This corresponds to an upper limit on the effective Majorana mass, $m_{\beta\beta} \simeq 28$--$128$ meV at 90\% CL~\cite{KamLAND-Zen:2024eml}.

A  major unresolved issue in neutrino physics is the origin of the observed flavor structure, in particular the pattern of neutrino masses and the large mixing angles in the lepton sector. One of the most widely explored approaches to this problem is based on non-Abelian discrete flavor symmetries, where leptons are assigned to specific representations of finite groups such as $A_4$, $S_4$, and related symmetries. These symmetries can impose non-trivial constraints on the structure of lepton mass matrices, leading to predictive relations among neutrino masses, mixing angles, and CP-violating phases. Consequently, discrete flavor symmetries have been extensively studied as a framework for understanding lepton flavor and neutrino mixing~\cite{Altarelli:2010gt,Ishimori:2010au,Chauhan:2023faf,Priya:2025khf}.

Despite their phenomenological successes, conventional discrete flavor models typically require the introduction of several flavon fields whose vacuum expectation values must be aligned in specific directions in order to reproduce the observed mixing pattern, particularly after the measurement of a nonzero reactor angle $\theta_{13}$. This requirement increases the complexity of the models and can reduce their predictive power. Motivated by these limitations, modular flavor symmetries have been proposed as an alternative framework in which Yukawa couplings transform non-trivially under the modular group and are expressed in terms of modular forms depending on a single complex modulus $\tau$~\cite{Feruglio:2017spp,Singh:2024imk,Kashav:2021zir,CentellesChulia:2023osj,Kumar:2023moh,Pathak:2025fpo}. In this approach, the flavor structure is largely determined by the modular symmetry itself, significantly reducing the number of free parameters and avoiding the need for flavon fields.

Most modular flavor models have been constructed within the context of low-energy supersymmetry, where holomorphic modular forms naturally arise in the superpotential. However, the continued absence of experimental evidence for supersymmetry at accessible energy scales has motivated the exploration of modular frameworks beyond supersymmetric constructions. In this regard, non-holomorphic modular symmetry provides a natural generalization of modular flavor models, allowing non-holomorphic modular forms while preserving modular invariance~\cite{Criado:2018thu,Novichkov:2023vay}. Such non-holomorphic modular frameworks do not rely on supersymmetry or holomorphy and retain the predictive power of modular symmetry. Recent studies have shown that these models can successfully accommodate current neutrino oscillation data, making them a well-motivated alternative for flavor model building in the absence of low-energy supersymmetry.

Non-holomorphic modular symmetry has been studied in various theoretical frameworks, including the Type-I seesaw mechanism~\cite{Nanda:2025lem}, Type-II seesaw mechanism~\cite{Nomura:2024atp}, Type-III seesaw mechanism~\cite{Priya:2025wdm}, Zee-Babu model~\cite{Nomura:2024nwh}, inverse seesaw mechanism~\cite{Zhang:2025dsa}, and related scenarios~\cite{Dey:2025zld,Kumar:2024uxn,Kumar:2025nut,Ding:2024inn,Nomura:2024vzw,Li:2024svh,Loualidi:2025tgw,Li:2025kcr,Okada:2025jjo,Abbas:2025nlv,Gao:2025jlw,Jangid:2025thp,Nomura:2025ovm,Nomura:2025raf, Nasri:2026nbf,Majhi:2026jdk, Tapender:2026ets,Nanda:2025fvw, Zhang:2026kyy}.

In this work, we explore the Type-I seesaw mechanism governed by three right-handed neutrinos transforming under a non-holomorphic modular symmetry. Within this framework, we perform a chi-square analysis to test the model against current experimental data. Future long-baseline neutrino oscillation experiments will provide further tests of the proposed framework.

The paper is organized as follows: the model framework and formalism are presented in Section~\ref{model_and_formalism}. The numerical results and their discussion are given in Section~\ref{Numerical Analysis and Discussion}, and finally, the conclusions are summarized in Section~\ref{Conclusions}.



\begin{table}[t]
\centering
\caption{Neutrino oscillation data used in the numerical analysis taken from NuFIT~6.0~\cite{Esteban:2024eli}.}
\label{tab:neutrino_data}
\renewcommand{\arraystretch}{1.2}
\begin{tabular}{lcc}
\hline\hline
\textbf{Parameter} & \textbf{NH} & \textbf{IH} \\
\hline
$\sin^2\theta_{12}$ 
& $0.308^{+0.012}_{-0.011}$ 
& $0.308^{+0.012}_{-0.011}$ \\

$\sin^2\theta_{23}$ 
& $0.470^{+0.017}_{-0.013}$ 
& $0.562^{+0.012}_{-0.015}$ \\

$\sin^2\theta_{13}$ 
& $0.02215^{+0.00056}_{-0.00058}$ 
& $0.02224^{+0.00056}_{-0.00057}$ \\

$\Delta m^2_{atm}\,[10^{-3}\,\mathrm{eV}^2]$ 
& $2.513^{+0.021}_{-0.019}$ 
& $-2.510^{+0.024}_{-0.025}$ \\

$\Delta m^2_{solar}\,[10^{-5}\,\mathrm{eV}^2]$ 
& $7.49^{+0.19}_{-0.19}$ 
& $7.49^{+0.19}_{-0.19}$ \\
\hline\hline
\end{tabular}
\end{table}

\section{Model} \label{model_and_formalism}
In this study, the particle content of the Standard Model is extended by three right-handed neutrino fields, which are singlets under the $SU(2)_L$ group and implement the Type-I seesaw mechanism. The right-handed neutrinos $N_1$, $N_2$, and $N_3$ transform as $1$, $1'$, and $1''$, respectively, under the modular $A_4$ symmetry, with modular weights assigned as $k_I = -2, 2,$ and $2$. The lepton doublets transform as $1$, $1''$, and $1'$ under $A_4$ with modular weight $k_I = 4$. The right-handed charged lepton singlets $e_R$, $\mu_R$, and $\tau_R$ are assigned as $1$, $1'$, and $1''$ representations of $A_4$, respectively, with modular weight $k_I = 2$. The modular weight assignments of the fields are chosen in such a way that they ensure modular invariance of the Yukawa terms and generate the required structure of the Dirac and Majorana mass matrices. The Higgs doublet is taken to be a singlet under the modular $A_4$ symmetry and is assigned zero modular weight. The charge assignments under the modular group $A_4$ and modular weights for the model are summarized in Table~\ref{tab:charges}. The modular-invariant Lagrangian is given by

\begin{equation}
\begin{split}
-\mathcal{L} = \; & \alpha \bar{L}_e H e_R Y_1^6 + \beta \bar{L}_\mu H \mu_R Y_1^6 + \gamma \bar{L}_\tau H \tau_R Y_1^6 \\
& + g_1 \bar{L}_e N_1 H + g_2 \bar{L}_\mu N_2 H Y_1^4 + g_2 \bar{L}_\tau N_3 Y_1^4 \\
& + g_3 \bar{L}_e N_3 Y_{1'}^4 + g_3 \bar{L}_\tau N_2 Y_{1'}^4 \\
& + M_0 N_1 N_1 Y_1^{-4} + M_1 (N_2 N_3 + N_3 N_2) Y_1^4 + M_2 N_2 N_2 Y_{1'}^4 \\
& + \text{H.c.}
\end{split}
\label{eq:Lagrangian}
\end{equation}

The charged lepton mass matrix is diagonal according to this model content which can be written as 

\begin{equation}
M_L = \frac{v}{\sqrt{2}}
\begin{pmatrix}
\alpha Y_1^6 & 0 & 0 \\
0 & \beta Y_1^6 & 0 \\
0 & 0 & \gamma Y_1^6
\end{pmatrix}
\label{eq:DiracMatrix}
\end{equation}
Here, $v$ is the vacuum expectation value(vev) of Higgs field. By choosing the real values of constants $\alpha, \beta~\text{and}~\gamma$, one can reproduce the correct values of lepton masses. The Dirac matrix according to this Lagrangian can be written as 
\begin{equation}
    M_D = \frac{v}{\sqrt{2}}
    \begin{pmatrix}
        g_1 & 0 & g_3Y_{1'}^4 \\
       0 & g_2 Y_1^4 & 0\\
       0 & g_3Y_{1'}^4 & g_2Y_1^4
    \end{pmatrix}
\end{equation}

The corresponding Majorana mass can be written as 
\begin{equation}
    M_N = 
    \begin{pmatrix}
        M_0 Y_1^{-4} & 0 & 0 \\
        0 & M_2 Y^4_{1'} & M_1 Y_1^4 \\
        0 & M_1 Y_1^4 & 0 \\
    \end{pmatrix}
\end{equation}

The Type-I seesaw mechanism is given as 
\begin{equation}
    m_\nu = M_D M_R^{-1}M_D^T.
\end{equation}

Finally the $m_\nu$ can be diagonalized as $U_{PMNS}^T M_\nu U_{PMNS} = \text{diag}(m_{\nu_1}, m_{\nu_2}, m_{\nu_3})$. The mixing angle now can be extracted from the mixing matrix $U_{PMNS}$ as

\begin{table}[t]
\caption{Field content and charge assignments of the Type-I seesaw mechanism under $A_4$ and modular weights.}
\centering
\small
\begin{tabular}{ccccc}
\hline\hline
 & $\bar{L}_e, \bar{L}_\mu, \bar{L}_\tau$ & $e_R, \mu_R, \tau_R$ & $N_1, N_2, N_3$ & $H$ \\
\hline
$SU(2)$ & 2 & 1 & 1 & 2 \\
$A_4$ & $1, 1'', 1'$ & $1, 1', 1''$ & $1, 1', 1''$ & 1 \\
$k_I$ & 4 & 2 & $-2, 2, 2$ & 0 \\
\hline\hline
\end{tabular}
\label{tab:charges}
\end{table}

\begin{equation}
    \sin^2 \theta_{13} = |U_{13}|^2, \quad
    \sin^2 \theta_{12} = \frac{|U_{12}|^2}{1 - |U_{13}|^2}, \quad
    \sin^2 \theta_{23} = \frac{|U_{23}|^2}{1 - |U_{13}|^2}.
\end{equation}
The Dirac $CP$-violating phase $\delta_{CP}$ can be determined from the PMNS matrix elements through the Jarlskog invariant, defined as
\begin{equation}
    J_{CP} = \text{Im}\left[U_{11} U_{22} U^{*}_{12} U^{*}_{21}\right] = s_{23} c_{23} s_{12} c_{12} s_{13} c_{13}^2 \sin\delta_{CP},
\end{equation}
where \( s_{ij} = \sin\theta_{ij} \) and \( c_{ij} = \cos\theta_{ij} \). In addition to $\delta_{CP}$, the Majorana $CP$ phases can be investigated using the PMNS matrix elements as
\begin{equation}
    I_1 = \text{Im}\left[U_{11}^* U_{12}\right] = c_{12} s_{12} c_{13}^2 \sin\left(\frac{\alpha_{21}}{2}\right),
\end{equation}
\begin{equation}
    I_2 = \text{Im}\left[U_{11}^* U_{13}\right] = c_{12} s_{13} c_{13} \sin\left(\frac{\alpha_{31}}{2} - \delta_{CP}\right).
\end{equation}
The effective Majorana mass($m_{\beta \beta}$) can be given as 
\begin{equation}
m_{\beta\beta} =
\left|
\begin{aligned}
& m_{1}\cos^{2}\theta_{12}\cos^{2}\theta_{13}
+ m_{2}\sin^{2}\theta_{12}\cos^{2}\theta_{13}\,e^{i\alpha_{21}} \\
& + m_{3}\sin^{2}\theta_{13}\,e^{i(\alpha_{31}-2\delta_{\rm CP})}
\end{aligned}
\right|.
\end{equation}

\section{Numerical Analysis and Discussion}
\label{Numerical Analysis and Discussion}
In this work, we have investigated the Type-I seesaw mechanism within the framework of non-holomorphic modular symmetry. To examine the consistency of the model with current experimental data, we have performed a chi-square analysis and identified a $\chi^2_{\min} = 7.06$, corresponding to the best-fit region of the parameter space. The chi-square function used in the analysis is defined as
\begin{equation}
\chi^2 = \sum_{i=1}^{5} \left( \frac{P_i - P_i^{0}}{\sigma_i} \right)^2,
\label{chisq}
\end{equation}

\begin{figure*}
\centering
\includegraphics[width=1\linewidth]{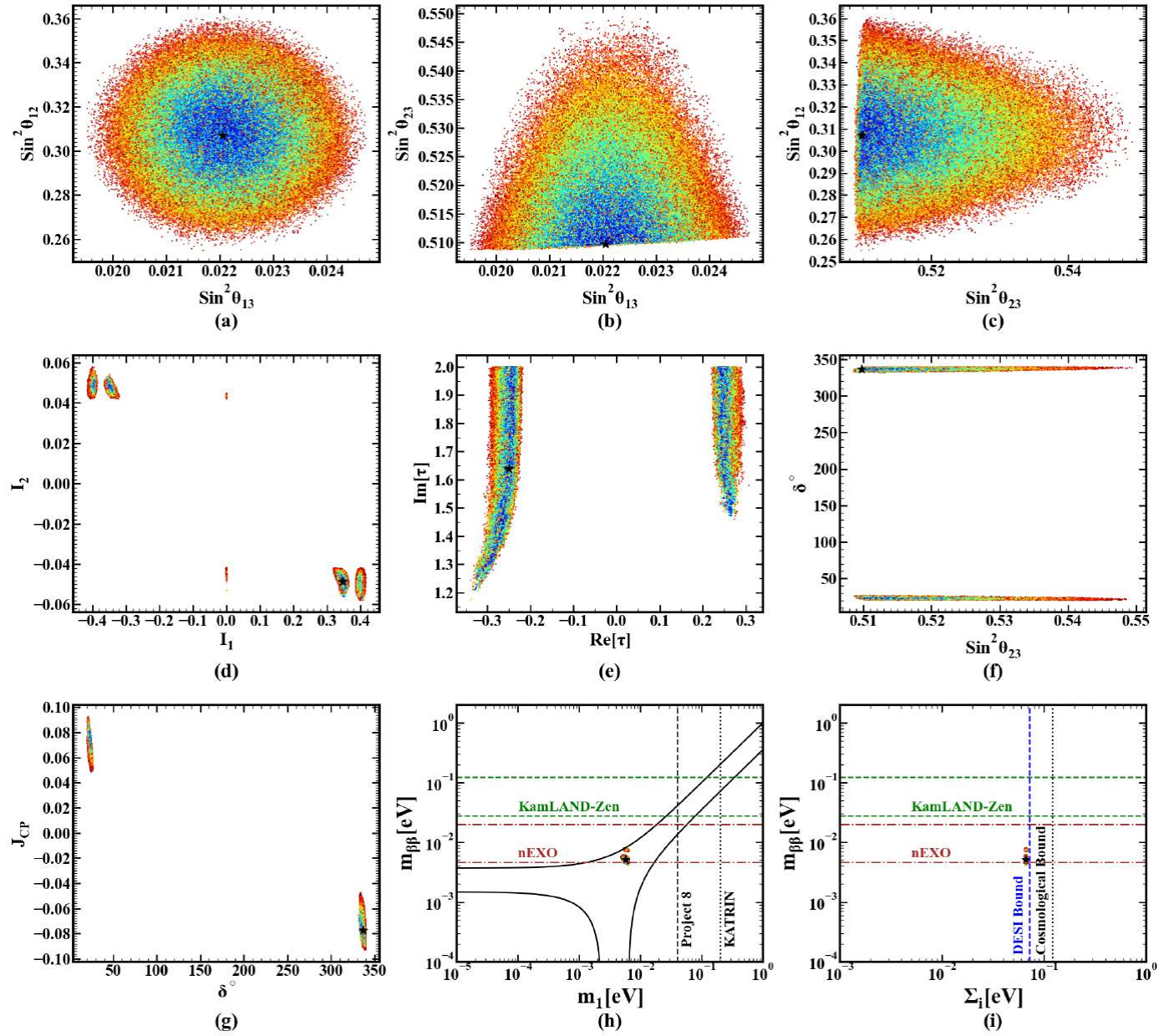}
\includegraphics[width=0.4\linewidth]{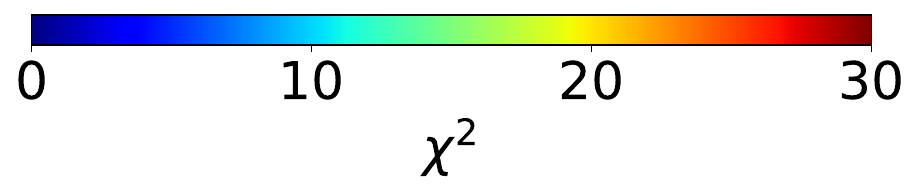}

\caption{Correlation plots among mixing angles, CP invariants ($I_1$, $I_2$), real and imaginary parts of the complex modulus $\tau$, Dirac-type CP phase ($\delta$) and atmospheric mixing angle ($\sin^2\theta_{23}$), Jarlskog invariant versus $\delta$, effective Majorana mass ($m_{\beta\beta}$) versus the lightest neutrino mass, and $m_{\beta\beta}$ versus the sum of neutrino masses ($\Sigma_i$). The solid star indicates the best-fit value corresponding to $\chi^2_{\rm min} = 7.06$. The color bar shows $\chi^2$ values from 0 to 30.}
\label{fig}
\end{figure*}

\begin{table}[t]
\caption{The input parameters of model with respect to $\chi^2_{min} = 7.06$.}
\centering
\small
\begin{tabular}{cccccc}
\hline\hline
$g_1$& $g_2$ & $g_3$ & $M_0 (GeV)$ & $M_1 (GeV)$ & $M_2 (GeV)$\\
\hline
0.0152 & 0.0311 & 0.0262 & $7.83 \times 10^{11}$ &$7.78 \times 10^{7}$ & $9.87 \times 10^{11}$ \\
\hline\hline
\end{tabular}
\label{tab:neutrino_params_2}
\end{table}

where, $P_i$ is the prediction of neutrino observable from the model, $P_i^0$ is best fit value from the Table~\ref{tab:neutrino_data} and $\sigma_i$ shows the uncertainty of neutrino observables at $1\sigma$ level.In this work, we have sampled five neutrino observables, i.e, three mixing angles($\sin^2\theta_{13}, \sin^2\theta_{12}, \sin^2\theta_{23}$) and two mass squared differences($\Delta m_{solar}^2$ and $\Delta m_{atm}^2$).  The input parameters used in the model are listed in Table~\ref{tab:neutrino_params_2}, corresponding to the minimum value $\chi^2_{\min} = 7.06$. Fig. \ref{fig}(a), \ref{fig}(b), \ref{fig}(c) shows that how consistent our model with the current experimentally predicted values of mixing angles. All five observables are consistent with the experimental data at the $1\sigma$ level, except for $\sin^2\theta_{23}$, which lies in the second octant. The best fit values of mixing angles with respect to $\chi_{min}^2 = 7.06$ are shown in Table~\ref{tab:neutrino_params}. In this work, the atmospheric mixing angle favors the second octant, with a best-fit value $\sin^2\theta_{23}=0.509$ correspond to the minimum $\chi^2$ at 7.06. This result is consistent with the indications from long-baseline experiments such as DUNE~\cite{DUNE:2020fgq}, NO$\nu$A~\cite{NOvA:2024}, and Hyper-Kamiokande~\cite{Hyper-Kamiokande:2018ofw}. The correlation between CP invariants ($I_1$ and $I_2$) is shown in Fig. \ref{fig}(d). The correlation between the real and imaginary parts of the complex modulus $\tau$ is presented in Fig.~\ref{fig}(e), with $\tau$ restricted to the fundamental domain. In our numerical scan, we consider the region $-0.5 \leq \mathrm{Re}(\tau) \leq 0.5$ and $0 < \mathrm{Im}(\tau) \leq 2$. The best-fit point obtained from the numerical analysis corresponds to the complex modulus $\tau = -0.24 + 1.63 i$. The correlation between the Dirac CP phase and the atmospheric mixing angle is shown in Fig.~\ref{fig}(f). The best-fit value of the Dirac phase is $\delta = 336.65^\circ$, lying in the fourth quadrant, and remaining compatible with current oscillation data at the $3\sigma$ level. In this framework, the only source of complex phases in the neutrino sector arises from the modular forms, which depend on the $\tau$. Since the phase structure of the modular forms is governed by a single complex parameter $\tau$, the available phase freedom in the neutrino mass matrix is limited. As a result, the Dirac CP phase $\delta_{CP}$ is constrained to a relatively narrow region of the parameter space, leading to comparatively small CP violation in our numerical results. This prediction can be tested in upcoming long-baseline measurements\cite{DUNE:2020fgq, NOvA:2024, Hyper-Kamiokande:2018ofw}. The best fit values of neutrino observables corresponding to $\chi^2_{min} = 7.06$ are shown in Table~\ref{tab:neutrino_params}. The correlation between the Jarlskog invariant and $\delta$ is shown in Fig.~\ref{fig}(g). The effective Majorana mass as a function of the lightest neutrino mass is also shown in Fig.~\ref{fig}(h). The effective Majorana mass is compatible with the expected sensitivity of the nEXO experiment~\cite{nEXO:2021ujk}. The predicted lightest mass also satisfies the bounds from KATRIN~\cite{KATRIN:2022ayy} and Project 8~\cite{Project8:2022wqh} experiments. Furthermore, the effective Majorana mass is plotted against the sum of neutrino masses in Fig~\ref{fig}(i), and the predicted sum is consistent with current cosmological constraints($\Sigma_i<0.12 eV$)~\cite{Planck:2018vyg} and bound proposed by DESI experiment($\Sigma_i<0.072 eV$)~\cite{DESI:2024mwx}. For the inverted hierarchy (IH) case, the minimum chi-square value is found to be very large, $\chi^2_{\min} = 221.6$. In this case, the predicted mixing angles $\sin^2\theta_{12}$ and $\sin^2\theta_{23}$ lie outside their allowed $3\sigma$ ranges, indicating a strong inconsistency with the current neutrino oscillation data. Consequently, the results for the IH case are not presented. For completeness, the corresponding neutrino observables associated with $\chi^2_{\min} = 221.6$ for IH are listed in Table~\ref{tab:neutrino_params_1}.

\begin{table}[t]
\caption{Best-fit values of neutrino oscillation parameters from the $\chi^2$ analysis for NH, with a minimum $\chi^2 = 7.06$.}
\centering
\small
\begin{tabular}{cccccc}
\hline\hline
$\sin^2\theta_{12}$ & $\sin^2\theta_{23}$ & $\sin^2\theta_{13}$ & $\delta_{CP} (^\circ)$ & $\Delta m^2_{21} (\text{eV}^2)$ & $\Delta m^2_{31} (\text{eV}^2)$ \\
\hline
0.307 & 0.509 & 0.022 & 336.65 & $7.5\times10^{-5}$ & $2.50\times10^{-3}$ \\
\hline\hline
\end{tabular}
\label{tab:neutrino_params}
\end{table}

\begin{table}[t]
\caption{Best-fit values of neutrino oscillation parameters from the $\chi^2$ analysis for IH, with a minimum $\chi^2 = 221.6$.}
\centering
\small
\begin{tabular}{cccccc}
\hline\hline
$\sin^2\theta_{12}$ & $\sin^2\theta_{23}$ & $\sin^2\theta_{13}$ & $\delta_{CP} (^\circ)$ & $\Delta m^2_{21} (\text{eV}^2)$ & $\Delta m^2_{31} (\text{eV}^2)$ \\
\hline
0.048 & 0.424 & 0.022 & 33.30 & $7.6\times10^{-5}$ & $-2.51\times10^{-3}$ \\
\hline\hline
\end{tabular}
\label{tab:neutrino_params_1}
\end{table}

\section{Conclusions}
\label{Conclusions}
In this work, we have studied neutrino mass generation within a Type-I seesaw framework based on non-holomorphic modular $A_4$ symmetry. The modular symmetry plays a central role in constraining the Yukawa sector, resulting in a predictive neutrino mass matrix with a reduced number of free parameters. Using a $\chi^2$ numerical analysis, our results show that the model provides an overall good description of the data for the normal hierarchy of neutrino masses, with the best-fit value obtained at $\chi^2_{min} = 7.06$. All neutrino oscillation parameters are found to lie within their $1\sigma$ allowed ranges, except for the atmospheric mixing angle $\theta_{23}$, which is predicted to lie in the second octant. The Dirac CP-violating phase is constrained to the first and fourth quadrants, indicating relatively weak CP violation. The effective Majorana neutrino mass is consistent with the projected sensitivity of nEXO experiment, while the predicted sum of neutrino masses remains consistent with present cosmological bounds and the stringent bound provided by the DESI experiment. For the NH, the values of various parameters predicted by the model are $\sin^2\theta_{13} = 0.022$, $\sin^2\theta_{12} = 0.307$, $\sin^2\theta_{23} = 0.509$, $\delta_{CP} = 336.65^\circ$, $m_1 = 0.0057eV$, $m_2 =0.0104eV $, $m_3 = 0.050eV$, $\Sigma_i = 0.066eV$, $I_1 = 0.346 $, $I_2 = -0.0486$, $J_{CP} = -0.0774 $, $x = -0.24$, $y = 1.63$. The inverted hierarchy (IH) is not viable in this model, as it predicts a large minimum value of $\chi^2$. Therefore, we do not discuss the IH case further. For completeness, the values predicted by our model for IH are listed in Table~\ref{tab:neutrino_params_1}.

Overall, the non-holomorphic modular invariant Type-I seesaw framework offers a coherent and predictive approach to neutrino masses and mixing. Future precision measurements, particularly from long-baseline neutrino oscillation experiments and neutrinoless double beta decay searches, will play a crucial role in testing the viability of this scenario.

\section*{Acknowledgments}
Priya and B. C. Chauhan wants to acknowledge the IUCAA for providing the HPC facility to carry out this work. Cheshta and Suneel Dutt acknowledge the support and facilities provided by the Department of Physics, Dr. B. R. Ambedkar National Institute of Technology (NIT) Jalandhar.


\begin{thebibliography}{100}

\bibitem{Super-Kamiokande:1998kpq}
Y.~Fukuda \textit{et al.} [Super-Kamiokande],
Phys.\ Rev.\ Lett.\ \textbf{81}, 1562 (1998).

\bibitem{SNO:2002tuh}
Q.~R.~Ahmad \textit{et al.} [SNO],
Phys.\ Rev.\ Lett.\ \textbf{89}, 011301 (2002).

\bibitem{KamLAND:2002uet}
K.~Eguchi \textit{et al.} [KamLAND],
Phys.\ Rev.\ Lett.\ \textbf{90}, 021802 (2003).

\bibitem{Minkowski:1977sc}
P.~Minkowski,
Phys.\ Lett.\ B \textbf{67}, 421 (1977).

\bibitem{Mohapatra:1979ia}
R.~N.~Mohapatra and G.~Senjanovic,
Phys.\ Rev.\ Lett.\ \textbf{44}, 912 (1980).

\bibitem{Foot:1988aq}
R.~Foot \textit{et al.},
Z.\ Phys.\ C \textbf{44}, 441 (1989).

\bibitem{Leitner:2017jco}
R.~Leitner [Daya Bay],
Nucl.\ Part.\ Phys.\ Proc.\ \textbf{285--286}, 32 (2017).

\bibitem{RENO:2012mkc}
J.~K.~Ahn \textit{et al.} [RENO],
Phys.\ Rev.\ Lett.\ \textbf{108}, 191802 (2012).

\bibitem{DoubleChooz:2011ymz}
Y.~Abe \textit{et al.} [Double Chooz],
Phys.\ Rev.\ Lett.\ \textbf{108}, 131801 (2012).

\bibitem{Esteban:2024eli}
I.~Esteban \textit{et al.},
JHEP \textbf{12}, 216 (2024).

\bibitem{Priya:2025avk}
Priya \textit{et al.},
Phys.\ Part.\ Nucl.\ Lett.\ \textbf{22}, 1227 (2025).

\bibitem{Ankush:2023pax}
Ankush \textit{et al.},
JCAP \textbf{08}, 062 (2023).

\bibitem{Schechter:1981bd}
J.~Schechter and J.~W.~F.~Valle,
Phys.\ Rev.\ D \textbf{25}, 2951 (1982).

\bibitem{Wolfenstein:1981rk}
L.~Wolfenstein,
Phys.\ Lett.\ B \textbf{107}, 77 (1981).

\bibitem{KamLAND-Zen:2024eml}
S.~Abe \textit{et al.} [KamLAND-Zen],
arXiv:2406.11438 [hep-ex] (2024).

\bibitem{nEXO:2021ujk}
G.~Adhikari \textit{et al.} [nEXO],
J.\ Phys.\ G \textbf{49}, 015104 (2022).

\bibitem{CUORE:2019yfd}
D.~Q.~Adams \textit{et al.} [CUORE],
Phys.\ Rev.\ Lett.\ \textbf{124}, 122501 (2020).

\bibitem{CUORE:2020boz}
A.~Campani \textit{et al.} [CUORE],
Int.\ J.\ Mod.\ Phys.\ A \textbf{35}, 2044016 (2020).

\bibitem{LEGEND:2017cdu}
N.~Abgrall \textit{et al.} [LEGEND],
AIP Conf.\ Proc.\ \textbf{1894}, 020027 (2017).

\bibitem{Altarelli:2010gt}
G.~Altarelli and F.~Feruglio,
Rev.\ Mod.\ Phys.\ \textbf{82}, 2701 (2010).

\bibitem{Ishimori:2010au}
H.~Ishimori \textit{et al.},
Prog.\ Theor.\ Phys.\ Suppl.\ \textbf{183}, 1 (2010).

\bibitem{Chauhan:2023faf}
G.~Chauhan \textit{et al.},
Prog.\ Part.\ Nucl.\ Phys.\ \textbf{138}, 104126 (2024).

\bibitem{Priya:2025khf}
Priya \textit{et al.},
arXiv:2501.00776 [hep-ph] (2025).

\bibitem{Feruglio:2017spp}
F.~Feruglio,
arXiv:1706.08749 [hep-ph].

\bibitem{Singh:2024imk}
L.~Singh, M.~Kashav and S.~Verma,
Nucl.\ Phys.\ B \textbf{1007}, 116666 (2024).

\bibitem{Kashav:2021zir}
M.~Kashav and S.~Verma,
JHEP \textbf{09}, 100 (2021).

\bibitem{CentellesChulia:2023osj}
S.~Centelles Chuli{\'a} \textit{et al.},
Phys.\ Rev.\ D \textbf{109}, 035016 (2024).

\bibitem{Kumar:2023moh}
R.~Kumar \textit{et al.},
Phys.\ Lett.\ B \textbf{853}, 138635 (2024).

\bibitem{Pathak:2025fpo}
G.~Pathak and M.~K.~Das,
arXiv:2508.13578 [hep-ph] (2025).

\bibitem{Criado:2018thu}
J.~C.~Criado and F.~Feruglio,
SciPost\ Phys.\ \textbf{5}, 042 (2018).

\bibitem{Novichkov:2023vay}
P.~P.~Novichkov \textit{et al.},
arXiv:2310.20681 [hep-ph] (2023).


\bibitem{Nanda:2025lem}
S.~K.~Nanda \textit{et al.},
arXiv:2509.22108 [hep-ph] (2025).

\bibitem{Nomura:2024atp}
T.~Nomura and H.~Okada,
Phys.\ Lett.\ B \textbf{868}, 139763 (2025).

\bibitem{Priya:2025wdm}
Priya \textit{et al.},
JHEP \textbf{01}, 036 (2026).


\bibitem{Nomura:2024nwh}
T.~Nomura and H.~Okada,
Phys.\ Lett.\ B \textbf{867}, 139618 (2025).

\bibitem{Zhang:2025dsa}
X.~Zhang and Y.~Reyimuaji,
Phys.\ Rev.\ D \textbf{112}, 075050 (2025).

\bibitem{Dey:2025zld}
M.~Dey,
arXiv:2509.10373 [hep-ph] (2025).

\bibitem{Kumar:2024uxn}
B.~Kumar and M.~K.~Das,
Int.\ J.\ Mod.\ Phys.\ A \textbf{40}, 2550090 (2025).

\bibitem{Kumar:2025nut}
B.~Kumar and M.~K.~Das,
arXiv:2509.01205 [hep-ph] (2025).

\bibitem{Ding:2024inn}
G.-J.~Ding \textit{et al.},
JHEP \textbf{01}, 191 (2025).

\bibitem{Nomura:2024vzw}
T.~Nomura \textit{et al.},
Phys.\ Lett.\ B \textbf{860}, 139171 (2025).

\bibitem{Li:2024svh}
C.-C.~Li \textit{et al.},
JHEP \textbf{12}, 189 (2024).

\bibitem{Loualidi:2025tgw}
M.~A.~Loualidi \textit{et al.},
Phys.\ Rev.\ D \textbf{112}, 015008 (2025).

\bibitem{Li:2025kcr}
C.-C.~Li and G.-J.~Ding,
arXiv:2509.15183 [hep-ph] (2025).

\bibitem{Okada:2025jjo}
H.~Okada and Y.~Orikasa,
arXiv:2501.15748 [hep-ph] (2025).

\bibitem{Abbas:2025nlv}
M.~Abbas,
PHEP \textbf{2025}, 7 (2025).

\bibitem{Gao:2025jlw}
X.-Y.~Gao and C.-C.~Li,
arXiv:2512.07158 [hep-ph] (2025).

\bibitem{Jangid:2025thp}
S.~Jangid and H.~Okada,
arXiv:2510.17292 [hep-ph] (2025).

\bibitem{Nomura:2025ovm}
T.~Nomura \textit{et al.},
JHEP \textbf{09}, 163 (2025).

\bibitem{Nomura:2025raf}
T.~Nomura and H.~Okada,
arXiv:2506.02639 [hep-ph] (2025).

\bibitem{Nasri:2026nbf}
S.~Nasri, L.~Singh, Tapender and S.~Verma,
[arXiv:2601.06435 [hep-ph]].

\bibitem{Majhi:2026jdk}
R.~Majhi, M.~K.~Behera and R.~Mohanta,
[arXiv:2602.23018 [hep-ph]].

\bibitem{Tapender:2026ets}
Tapender and S.~Verma,
[arXiv:2602.17243 [hep-ph]].

\bibitem{Nanda:2025fvw}
S.~K.~Nanda, M.~Ricky Devi, \textit{et al.},
[arXiv:2512.24132 [hep-ph]].

\bibitem{Zhang:2026kyy}
X.~Zhang and Y.~Reyimuaji,
[arXiv:2603.19104 [hep-ph]].

\bibitem{DUNE:2020fgq}
B.~Abi \textit{et al.} [DUNE],
Eur.\ Phys.\ J.\ C \textbf{81}, 322 (2021).

\bibitem{NOvA:2024}
I.~Singh, B.~C.~Choudhary and L.~Suter [NOvA],
FERMILAB-CONF-24-0792-PPD (2024).

\bibitem{Hyper-Kamiokande:2018ofw}
K.~Abe \textit{et al.} [Hyper-Kamiokande],
[arXiv:1805.04163 [physics.ins-det]].

\bibitem{KATRIN:2022ayy}
M.~Aker \textit{et al.} [KATRIN],
J.\ Phys.\ G \textbf{49}, 100501 (2022).

\bibitem{Project8:2022wqh}
A.~A.~Esfahani \textit{et al.} [Project~8],
arXiv:2203.07349 [nucl-ex] (2022).

\bibitem{Planck:2018vyg}
N.~Aghanim \textit{et al.} [Planck],
Astron.\ Astrophys.\ \textbf{641}, A6 (2020).

\bibitem{DESI:2024mwx}
A.~G.~Adame \textit{et al.} [DESI],
JCAP \textbf{02}, 021 (2025).

\end{thebibliography}

\end{document}